\documentclass[aps,pre,preprint,superscriptaddress,amsmath,amssymb,nofootinbib]{revtex4}
\usepackage{amsmath}
\usepackage{graphicx}
\usepackage{amssymb}
\usepackage{dcolumn}
\usepackage{bm}
\usepackage{color}
\usepackage{float}
\usepackage{upgreek}
\usepackage[final]{hyperref} 
\hypersetup{
	colorlinks=true,       
	linkcolor=blue,        
	citecolor=blue,        
	filecolor=magenta,     
	urlcolor=blue         
}
\usepackage{upgreek}
\usepackage[normalem]{ulem}

\makeatletter

\begin{document}

\title{Mass separation in an asymmetric channel}

\author{Narender Khatri}\thanks{narenderkhatri8@iitkgp.ac.in}
\affiliation{Department of Physics, Indian Institute of Technology Kharagpur, Kharagpur - 721302, India}
\author{P. S. Burada}\thanks{Corresponding author: psburada@phy.iitkgp.ac.in}
\affiliation{Department of Physics, Indian Institute of Technology Kharagpur, Kharagpur - 721302, India}

\date{\today}
	
\begin{abstract} 
We present a mechanism to sort out particles of different masses in an asymmetric channel, where the entropic barriers arise naturally and control the diffusion of these particles. 
When particles are subjected to an oscillatory force, with the scaled amplitude $a$ and frequency $\omega$, the mean particle velocity exhibits a bell-shaped behavior as a function of the particle mass, indicating that particles with an optimal mass $m_{op}$ drift faster than other particles. 
By tuning $a$ and $\omega$, we get an empirical relation to estimate $m_{op} \sim (a\,\omega^2)^{-0.4}$. 
An additional static bias, applied in the opposite direction of the rectified velocity, would push the particles of lighter mass to move in its direction while the others drift opposite to it.
This study is useful to design lab-on-a-chip devices for separating particles of different masses.
\end{abstract}

%

\maketitle

\section{Introduction}

In many natural systems and industrial productions, matter consists of mixtures of various small parts on a range from macroscale to nanoscale, e.g., mixed polar and nonpolar macroparticles \cite{Brunet_JCP}, mixed bacteria \cite{Petrides}, mixed nanoparticles \cite{Ploschner_NL}, various DNA fragments \cite{Dorfman}, different kinds of cells \cite{Franke_BMF}, or size-dispersed particles in the liquid \cite{Marchesoni}, to name but a few. 
The separation and sorting of small-sized particles have a wide range of industrial, biomedical and clinical applications, e.g., wastewater purification, blood sample preparation, and disease diagnostic \cite{Jin}.
Various particle separation mechanisms or tools are in use to separate the desired particles from the bulk; for instance, templates consist of arrays of obstacles \cite{Reimann_prl}, centrifuge machines \cite{Lee_prl}, sieve or membrane \cite{Duke_prl}, and by applying some external fields \cite{Dholakia_Nature}.
These separation techniques are based on the drift or diffusive behavior of the particles whose properties depend on their mass, size, shape, and charge.
In particular, compared to the macro-sized particles, the controlled separation of the mesoscopic particles from the mixture, based on their physical properties, is in high demand for laboratory research and industrial applications \cite{Petrides, Dorfman, Marchesoni, Austin, Reguera_prl, Reimann_prl,Rubi_prl, Mukhopadhyay_prl, Sens_prl, Slapik_prl, PhysicaA, Bader_PNAS}.
Note that the ubiquitous Brownian dynamics is relevant particularly for 
the particles in the mesoscopic size regime.
The separation of mesoscopic constituents presents major challenges in various disciplines, from biomedical problems such as separating malignant circulating tumor cells from leukocytes in the bloodstream \cite{Jin} to technological problems on colloidal scales such as separating nanoparticles based on their physical properties \cite{Reguera_prl, Slapik_prl}. 
Remarkably, the effective isolation and separating techniques carry a strong potential to achieve selective transport of biological particles such as cells, organelles, or DNA complexes \cite{Bhagat_MBEC, Aksimentiev_BJ}.

Quite often, the separation of particles happens in microstructures such as porous media \cite{Schwartz_Nano}, microfluidic channels \cite{Muller}, and living tissues \cite{Newby_R}. 
These systems have attracted the attention of physicists, chemists, mathematicians, biologists, and engineers because the irregular shapes of the walls control the volume of the phase space accessible to the particles, due to which the entropic barriers arise
and influence the diffusion of the particles in such systems \cite{Reguera_prl, Burada_CPC, Burada_prl}. 
So far, much effort has been devoted to developing the size-based separation of particles \cite{Reguera_prl,Rubi_prl,Sens_prl,Slapik_pra,Karimi_pre}. 
However, needless to say, the effective isolation of selected particles from the mixture, based on their mass, is much in demand in chemistry, biology, nanotechnology, and industry \cite{Mukhopadhyay_prl,Slapik_prl,Nagaoka_pre,Borromeo_prl}.
It is because anomalies in bioparticles mass might be an important factor associated with disease initiation and propagation.
For instance, it has been observed that the mass of a cancer cell is higher than a healthy one \cite{Suresh}.
Thus, one needs to develop new innovative mass-based isolation techniques for the diagnosis and detection of diseases.
Note that this is a very challenging task because, typically, identically sized particles may have different masses \cite{Slapik_prl}.

In the present work, we introduce a mechanism to separate particles based on their mass when they diffuse in an asymmetric triangular channel under the influence of an oscillatory force and a static bias. 
The working principle relies on the entropic rectification of the noisy motion of particles in the presence of an oscillatory force \cite{Marchesoni, Reguera_prl, Muller, Reguera_pre, Khatri_JSTAT} and a static bias that further controls the movement of the particles.
In particular, the strength of the entropic rectification depends strongly on the mass of the particles.
Therefore, in the presence of a static bias which is applied in the opposite direction of the rectified velocity, it is possible
to separate particles of lighter mass to move in its direction while the others drift opposite to it.
The lighter particles follow the static bias, whereas heavier particles drift in the opposite direction resulting in an efficient and faster particle separation.  

The rest of this article is organized in the following way. 
In Sec.~\ref{sec:Model}, we introduce our model for the inertial Brownian particles of different masses subjected to external forces in a two-dimensional asymmetric channel. 
The results in the absence and presence of a static bias are discussed in Sec.~\ref{sec:Ab_staticbias} and Sec.~\ref{sec:staticbias}, respectively.
Finally, we present the main conclusions in Sec.~\ref{Conclusions}.

\section{Model}
\label{sec:Model}

\begin{figure}[H]
\centering
\includegraphics[scale = 1.3]{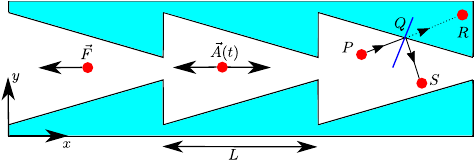}
\caption{Schematic illustration of a two-dimensional triangular channel with a periodicity $L$ confining the motion of a Brownian particle of mass $M$. The structure is defined by Eq.~(\ref{eq:wall}). The particle is driven by a static bias $\vec{F}$ and an oscillatory force $\vec{A}(t)$ along the $x-$direction.  
}
\label{fig:chw}
\end{figure}

In order to illustrate the mass-based separation mechanism, we have chosen a triangular channel depicted in Fig.~\ref{fig:chw}.
A Brownian particle of mass $M$ is diffusing in this channel, filled with a fluid of viscosity $\mu$, subjected to an unbiased oscillatory force $\vec{A}(t)$ and 
a static bias $\vec{F}$; both are acting along the principal axis of the channel. 
The Brownian dynamics of such a particle is represented by the Langevin equation
\begin{equation}
M \frac{d^2 \vec{r}}{d t^2} = -\eta \frac{d \vec{r}}{d t} - [F + A(t)] \hat{x} +  \sqrt{\eta k_B T} \vec{\xi}(t),
\label{eq:Langevin1}  
\end{equation}
where $\vec{r}$ is the position of the particle, $\eta$ is the friction coefficient, $k_B$ is the Boltzmann constant, and $T$ is the temperature. 
The explicit form of the oscillatory force is given by $ A(t) = A \sin(\Omega t)$, where $A$ is the amplitude and $\Omega$ is the frequency of the sinusoidal driving. 
The Gaussian random force $\vec{\xi}(t)$ with zero mean is uncorrelated in time and therefore obeys the fluctuation-dissipation relation $\langle \xi_i (t)  \xi_j (t') \rangle = 2 \delta_{ij} \delta(t-t')$ for $i, j = x, y$.  
The shape of the upper wall is defined by 
\begin{equation}
w_u (x) = \begin{cases}
w_\mathrm{min}, & x = 0,\\
w_\mathrm{max} - (w_\mathrm{max} - w_\mathrm{min}) \frac{x}{L}, & 0 < x\leq L,
\end{cases}
\label{eq:wall}
\end{equation}
where $w_\mathrm{max}$ and $w_\mathrm{min}$ correspond to the maximum half-width and minimum half-width of the channel, respectively, and $L$ corresponds to the periodicity of the channel. 
The ratio of these two widths defines the dimensionless aspect ratio given by $\epsilon = w_\mathrm{min}/w_\mathrm{max}$.
Due to the symmetry about the principal axis of the channel, the lower wall is given by $w_l (x) = - w_u (x)$. 
Consequently, $2 w (x) = w_u (x) - w_l (x)$ corresponds to the local width of the channel.

In order to achieve a dimensionless description, we henceforth scale all lengths by the periodicity of the channel $L$ and time by $\tau = \eta L^2/(k_B T)$, which is the characteristic diffusion time \cite{Burada_prl, Khatri_pre}. 
In dimensionless form, the two-dimensional Langevin equation~(\ref{eq:Langevin1}) reads
\begin{equation}
m \frac{d^2 \vec{r}}{d t^2} = -\frac{d \vec{r}}{d t} - [f + a \sin(\omega t)] \hat{x} + \vec{\xi}(t).
\label{eq:Langevin2}  
\end{equation}
Here, the dimensionless mass is given by $m  = \tau_0/\tau = M k_B T/(\eta^2 L^2)$, where $\tau_0 = M/\eta$ defines the characteristic time of velocity relaxation for the free Brownian particle.
Therefore, the dimensionless mass $m$ depends not only on the actual physical mass of the particle $M$ but also on the friction coefficient $\eta$, thermal energy $k_B T$, and the periodicity of the channel $L$. 
The dimensionless static force reads $f = FL/(k_B T)$, which is the ratio of work done to the particle due to the external static force and the available thermal energy. 
Other parameters read $a = AL/(k_B T)$ and $\omega = \Omega \tau$. 
In the rest of the article, we use dimensionless variables only.

Since the particles along the $y$-direction are confined, the observable of foremost interest for mass separation is the stationary average velocity $\langle v \rangle$ of the particles along the $x$-direction.
Unfortunately, for the considered system, the Fokker-Planck equation corresponding to the Langevin equation~(\ref{eq:Langevin2}) cannot be solved analytically for $\langle v \rangle$ by any known analytical method \cite{Burada_prl}. 
Thus, $\langle v \rangle$ is calculated using Brownian dynamics simulations performed by solving the Langevin equation~(\ref{eq:Langevin2}) using the standard stochastic Euler algorithm over $10^4$ trajectories with reflecting boundary conditions at the channel walls to ensure the confinement within the channel.
Figure~\ref{fig:chw} shows the reflection of a particle at the channel boundary. 
Let us say initially the particle was at $P(x_1, y_1)$, and the position of the particle in the next instant of time is $R(x_2, y_2)$, which is outside of the channel boundary. 
The line joining points $P$ and $R$ intersects the channel boundary at a point $Q(\alpha, \beta)$, i.e., the reflection point which can be calculated numerically using the bisection method. 
The desired position $S(x_3, y_3)$ of the particle after reflection at point $Q$ is given by \cite{Khatri_JCP} 
\begin{subequations}
	\begin{alignat}{2}
	x_3 &= \alpha \pm \, \frac{l}{\sqrt{1 + m_3^2}},\\
	y_3 &= \beta + m_3(x_3 - \alpha).
	\end{alignat}
	\label{eq:fxy}
\end{subequations} 
Here, $m_3$ is the slope of line $QS$ and $l = \sqrt{(y_2 - \beta)^2 + (x_2 - \alpha)^2}$.
Note that depending on the driving force acting on the particle, the particle may reflect multiple times at the channel boundaries to reach the final position.
The average velocity of the particles, having mass $m$, along the $x$-direction is calculated, in the long time limit, as \cite{Reguera_prl, Khatri_JSTAT}
\begin{equation}
\langle v \rangle = \lim_{t\to\infty} \frac{\langle x(t) \rangle}{t}.
\label{eq:velocity}  
\end{equation}

\section{In the absence of static bias}
\label{sec:Ab_staticbias}

\begin{figure}[t]
	\centering
	\includegraphics{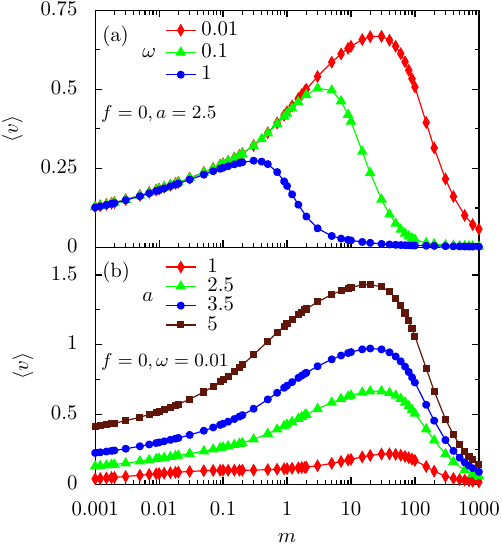}
	\caption{The average velocity $\langle v \rangle$ versus the mass $m$ of particles for various values of the driving frequency $\omega$ (a) and driving amplitude $a$ (b) in the absence of static force. The other parameters of the channel are $L = 1$, $w_\mathrm{max} = 1$, and $\epsilon = 0.1$.}
	\label{fig:graph2}
\end{figure}

The average velocity as a function of the mass of particles is depicted in Fig.~\ref{fig:graph2} for various values of the driving frequency $\omega$ and driving amplitude $a$ in the absence of static force.
We observe that particles of various masses exhibit rectification in the positive $x-$axis.
It is due to the fact that the oscillatory force can break the thermodynamic equilibrium and induces the directed transport because of the asymmetry of the channel.
Here, the positive velocity is due to the chosen shape of the channel structure. 
For example, if the channel shape is inverted with respect to the $y-$axis, the rectification will be in the negative $x-$axis.
Interestingly, the average velocity exhibits a peak at an optimal mass for which the particles show higher rectification compared to other particles.
The peak is more pronounced as $\omega$ decreases or $a$ increases.
Note that the optimal mass at which the maximum occurs can be effectively controlled by suitably tuning $\omega$ and $a$. 
When the particles are very heavy, the inertia starts to dominate over the strength of the oscillatory force; therefore, as one would expect, the velocity tends to zero.
In the other limit, i.e., $m \to 0$, $\langle v \rangle$ increases linearly with $m$, which is independent for different values of $\omega$. In this limit, as $m$ increases, due to the rise in the inertia, the effective drag from the surrounding fluid medium on the particles reduces. 
This leads to an increase in $\langle v \rangle$ (see Fig.~\ref{fig:graph2}(a)).
However, in the limit $\omega \rightarrow \infty$, $\langle v \rangle$ will tend to zero \cite{Khatri_JSTAT}.
It is because, in this limit, particles experience a zero time-averaged constant force $a = \int_{0}^{\frac{2 \pi}{\omega}} a(t) dt = 0$.
Also, in $a \to \infty$ limit, $\langle v \rangle$ will tend to zero because the effect of asymmetry of the channel will disappear in this limit \cite{Reguera_prl, Khatri_JSTAT}.
In particular, this asymptotic regime of very large $\omega$ and $a$ is not yet reached in Fig.~\ref{fig:graph2}.

\subsection{Influence of channel aspect ratio $\epsilon$}
\label{sec:channel_aspect}

\begin{figure}[htb]
\centering
\includegraphics[scale = 0.85]{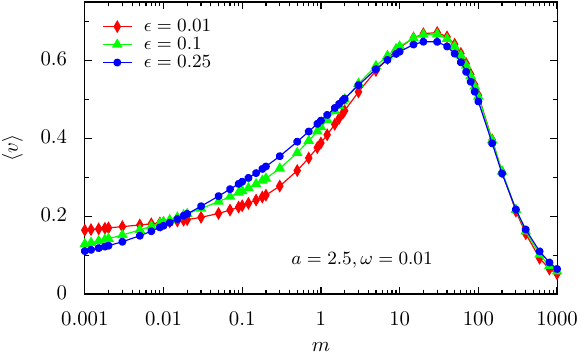}
\caption{The average velocity $\langle v \rangle$ versus the mass $m$ of particles for different values of $\epsilon$. The set parameters are $a = 2.5, \omega = 0.01, L = 1$, and $w_\mathrm{max} = 1$.}
\label{fig:graph3}
\end{figure}

In order to obtain the influence of channel aspect ratio $\epsilon$, in Fig.~\ref{fig:graph3}, we study the average velocity as a function of the mass of particles for different values of $\epsilon$.
It is evident that the bottleneck width of the channel does not influence the qualitative behavior of $\langle v \rangle$ as long as the condition $w_\mathrm{min} \ll w_\mathrm{max}$, i.e., $\epsilon \ll 1$, is satisfied. 
For $\epsilon = 1$, i.e., a flat channel, as one would expect, there would be no rectification of particles resulting in $\langle v \rangle = 0$.
Note that the main reason for the observed bell-shaped behavior of $\langle v \rangle$ is the asymmetry of the channel geometry.
In the $m \to 0$ limit, $\langle v \rangle$ decreases with $\epsilon$.
In the other limit, i.e., when the particles are very heavy, as mentioned earlier, $\langle v \rangle$ tends to zero irrespective of $\epsilon$.

\begin{figure}[t]
\centering
\includegraphics[scale = 1.8]{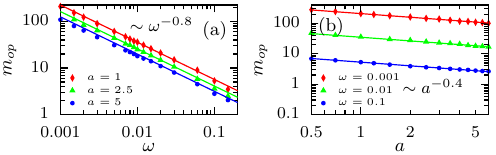}
\caption{Dependence of the optimal mass $m_{op}$ on the driving frequency $\omega$ (a) and driving amplitude $a$ (b). 
The lines correspond to the empirical relation $m_{op} \sim (a\,\omega^2)^{-0.4}$, see Table~\ref{table:constants}.
The other parameters of the channel are $L = 1$, $w_\mathrm{max} = 1$, and $\epsilon = 0.1$.}
\label{fig:graph4}
\end{figure}
\begin{table}
\begin{tabular}{ |p{1.5cm}|p{1.5cm}|p{1.5cm}|p{1.5cm}|}
\hline
\multicolumn{2}{|c|}{$m_{op} = k_1\,\omega^{-0.8}$}&\multicolumn{2}{c|}{$m_{op} = k_2\,a^{-0.4}$} \\[1.5ex]
\hline
$a$ & $k_1$ & $\omega$ & $k_2$\\[1.5ex] 
\hline
1   & 0.9   & 0.001 & 210   \\[1ex] 
2.5 & 0.65  & 0.01  & 35    \\[1ex]
5   & 0.5   & 0.1   & 5.3  \\[0.5ex]
\hline
\end{tabular}
\caption{Prefactors $k_1$ and $k_2$ correspond to the empirical relations for different values of $a$ and $\omega$, respectively.}
\label{table:constants} 
\end{table}

To get more insights into the optimal mass $m_{op}$, in Fig.~\ref{fig:graph4}, we extract $m_{op}$ for various values of the driving amplitude and frequency. 
The optimal mass $m_{op}$ decreases with $\omega$, for a fixed $a$ value, 
as $m_{op} = k_1\, \omega^{-0.8}$. 
Also, $m_{op}$ decreases with $a$, for a fixed $\omega$ value, 
as $m_{op} = k_2\, a^{-0.4}$. 
The Prefactors $k_1$ and $k_2$ are listed in table~\ref{table:constants} for a few values of $a$ and $\omega$, respectively.
Therefore, one can estimate that the value of the optimal mass varies as 
$m_{op} = k\, (a\,\omega^2)^{-0.4}$, where the prefactor $k \approx 0.91$. 
Here, $k$ is calculated by substituting the known values of $m_{op}$ corresponding to different $a$ and $\omega$ in the previous empirical relation. 
Thus, by tuning $a$ and $\omega$, the particles with an optimal mass $m_{op}$ can be rectified with higher speed compared to other particles. 
Note that in real units, the $k$ value (or $k_1, k_2$) depends on various parameters like $L$, $k_\mathrm{B}\,T$, and $\eta$. 
However, we have observed that the minimum half-width of the channel $w_\mathrm{min}$ (see Fig.~\ref{fig:chw}) does not influence the empirical relation for $m_{op}$ as long as the condition $\epsilon \ll 1$ is satisfied.

\section{Influence of the static bias}
\label{sec:staticbias}

\begin{figure}[htb]
	\centering
	\includegraphics[scale = 1]{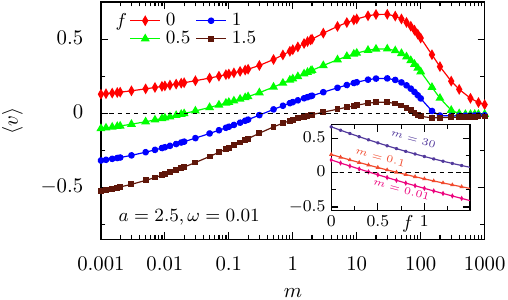}
	\caption{The average velocity $\langle v \rangle$ versus the mass $m$ of particles for different values of the static force $f$. 
		The inset depicts $\langle v \rangle$ as a function of $f$ for the particles of different masses. The set parameters are $a = 2.5$, $\omega = 0.01$, $L = 1$, $w_\mathrm{max} = 1$, and $\epsilon = 0.1$.}
	\label{fig:graph5}
\end{figure}

In order to achieve the separation of particles of different masses in opposite directions, a static force $f$ can be applied to control the transport direction. 
Figure~\ref{fig:graph5} depicts the average velocity dependence on the mass of particles for different strengths of the static bias for fixed values of $a$ and $\omega$ of the oscillatory force. 
In the absence of a static bias, due to the chosen channel structure, particles 
exhibit a net positive velocity. 
However, if we switch on a small static bias in the negative $x-$axis (see Fig.~\ref{fig:chw}), particles of the lighter mass move to the left side of the channel, whereas heavier particles drift to the right side (see Fig.~\ref{fig:graph5}). 
This mechanism provides a way to separate the lighter particles from the heavier ones. 
A similar behavior has been reported for the overdamped Brownian particles of different radii \cite{Reguera_prl}, i.e., large particles move towards the right side of the channel whereas small particles follow the static force. 
Note that the optimal mass $(m_{op})$ is independent of the static bias.
The inset of Fig.~\ref{fig:graph5} depicts the average velocity dependence on the static force for the particles of different masses. 
In particular, by tuning the static bias $f$, one can control the separation of particles of different masses. 
For example, by choosing $f = 1$, the particles of lighter mass $m = 0.01$ move towards the left side of the channel with speed $0.23$, whereas the particles of heavier mass $m = 30$ drift to the right side with a speed of $0.24$. 
Interestingly, as depicted in the inset of Fig.~\ref{fig:graph5},
the velocities are almost parallel and vary linearly with the strength of the static bias, thus facilitating an efficient control of the mass separation effect.

\begin{figure}[htb]
\centering
\includegraphics[scale = 1]{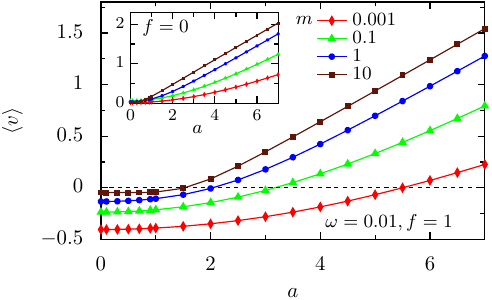}
\caption{The average velocity $\langle v \rangle$ versus the driving amplitude $a$ of the oscillatory force for the particles of different masses in the presence of static bias. 
The inset depicts the same in the absence of static bias. 
The set parameters are $\omega = 0.01$, $L = 1$, $w_\mathrm{max} = 1$, and $\epsilon = 0.1$.}
\label{fig:graph6}
\end{figure}

Figure~\ref{fig:graph6} shows the dependence of the average velocity $\langle v \rangle$ on the amplitude $a$ of the oscillatory force for the particles of different masses.  
As mentioned earlier, in the absence of the static force, particles of different masses move towards the positive $x-$axis, exhibiting a positive velocity (see inset of Fig.~\ref{fig:graph6}). 
However, in the presence of static bias, by suitably tuning $a$, one can control the separation of particles of different masses. 
For example, by keeping $a = 3$, the particles of mass $m < 0.1$ move to the left side of the channel, whereas particles of mass $m > 0.1$ drift to the right side. Moreover, the crossover of average velocity from negative to positive can also be controlled by $a$.

The observed mass-based separation mechanism can be studied experimentally for Brownian particles of various masses diffusing in an asymmetric channel, which can be prepared by microprinting on a substrate \cite{Maier_nlett, Mahmud_nature}.
The fabrication of identical particles of different masses is already within reach of today's nanotechnology \cite{Marchesoni, Muller}.
These particles can be prepared from hydrophobic materials such as luminescent polystyrene, polycrystalline silver, gold, copper, etc.
In order to have an estimate in real units, which is very useful for the experimentalists, the characteristic values of the friction coefficient and characteristic diffusion time for the particles in water moving in a triangular channel with aspect ratio $\epsilon = 0.1$ and period length $L \sim 10 ~\mu \mathrm{m}$ at room temperature ($T \sim 300 ~\mathrm{K}$) are given by $\eta \sim 2 ~ \mathrm{n\/kg/s}$ \cite{Cussler} and $\tau \sim 50~ \mathrm{s}$, respectively \cite{Marchesoni, Muller}.
In the absence of static force, for the parameters $A \sim 0.25 ~\mathrm{f\/N}$ and $\Omega \sim 0.0002 ~ \mathrm{s^{-1}}$, the particles of mass $M \sim 0.25  ~ \mathrm{nkg}$ move with a higher average velocity $\langle V \rangle \sim 0.13 ~ \mu\mathrm{m/s}$ than the other particles. 
Whereas, for the parameters $A \sim 0.25  ~ \mathrm{f\/N}$ and $\Omega \sim 0.002 ~ \mathrm{s^{-1}}$, the optimal mass is $M \sim 0.04  ~ \mathrm{n\/kg}$.
This suggests that by suitably tuning the driving frequency and amplitude, the particles of an optimal mass move faster than other particles in the same direction.
On the other hand, in the presence of a static bias, for the parameters $F \sim 0.1 ~ \mathrm{f\/N}$, $A \sim 0.25 ~ \mathrm{f\/N}$, and $\Omega \sim 0.0002 ~ \mathrm{s^{-1}}$, the particles of mass $M \sim 0.1 ~ \mathrm{n\/kg}$ move in the negative direction with an average velocity $|\langle V \rangle| \sim 0.06 ~ \mu\mathrm{m/s}$, and particles of mass $M \sim 0.3  ~ \mathrm{n\/kg}$ drift in the positive direction with the average velocity  $\langle V \rangle \sim 0.05 ~ \mu\mathrm{m/s}$. 
Therefore, in the presence of static force, the particles of different masses can be separated in opposite directions.  
It is expected that these results will motivate the experimentalists to design lab-on-a-chip devices for separating nano- and microparticles, proteins, organelles, and cells based on their mass.

\section{Conclusions}\label{Conclusions}

In this work, we have presented a mechanism for separating particles in an asymmetric channel based on their mass.  
This mechanism is based on the entropic rectification of the noisy motion of particles in the presence of an unbiased oscillatory force and a static force that controls the drift direction.
It is found that entropic rectification depends strongly on mass of the particles. 
We could demonstrate that the average velocity exhibits a maximum at an optimal mass for various parameters, and this optimal mass depends on the driving frequency and amplitude of the oscillatory force as $m_{op} \sim (a\,\omega^2)^{-0.4}$.  
In the presence of a small static force in the negative $x$-axis, we have found that lighter particles can be separated from the heavier ones by pushing them in opposite directions.
It is conceivable that this separation mechanism could be implemented in an asymmetric structure where the entropic effects are prominent to isolate nano- and microparticles, proteins, organelles, and cells.

\section{Acknowledge}

This work was supported by the Indian Institute of Technology
Kharagpur under the Grant No. IIT/SRIC/PHY/TAB/2015-16/114.


\begin{thebibliography}{100}
	
\bibitem{Brunet_JCP}
{C. Brunet, J. G. Malherbe, and S. Amokrane}, {J. Chem. Phys.} {\bf 130}, {134908} {(2009)}.	
	
\bibitem{Petrides}
{R. G. Harrison, P. W. Todd, S. R. Rudge, and D. P. Petrides}, {Bioseparations Science and Engineering} {(Oxford University Press, Oxford, 2003)}.	

\bibitem{Ploschner_NL}
{M. Ploschner, T. Cizmar, M. Mazilu, A. D. Falco, and K. Dholakia}, {Nano Lett.} {\bf 12}, {1923} {(2012)}.

\bibitem{Dorfman}
{K. D. Dorfman}, {Rev. Mod. Phys.} {\bf 82}, {2903} {(2010)}. 


\bibitem{Franke_BMF}
{T. M. Geislinger and T. Franke}, {Biomicrofluidics} {\bf 7}, {044120} {(2013)}. 



\bibitem{Marchesoni}
{P. H\"anggi and F. Marchesoni}, {Rev. Mod. Phys.} {\bf 81}, {387} {(2009)}. 

\bibitem{Jin}
{C. Jin, S. M. McFaul, S. P. Duffy, X. Deng, P. Tavassoli, P. C. Black, and  H. Ma}, {Lab Chip} {\bf 14}, {32} {(2014)}. 

\bibitem{Reimann_prl}
{L. Bogunovic, M. Fliedner, R. Eichhorn, S. Wegener, J. Regtmeier, D. Anselmetti, and P. Reimann}, {Phys. Rev. Lett.} {\bf 109}, {100603} {(2012)}.

\bibitem{Lee_prl}
{J. Lee and A. J. C. Ladd}, {Phys. Rev. Lett.} {\bf 89}, {104301} {(2002)}. 

\bibitem{Duke_prl}
{T. A. J. Duke and R. H. Austin}, {Phys. Rev. Lett.} {\bf 80}, {1552} {(1998)}.

\bibitem{Dholakia_Nature}
{M. P. MacDonald, G. C. Spalding, and K. Dholakia}, {Nature} {\bf 426}, {421} {(2003)}.


\bibitem{Austin}
{W. D. Volkmuth and R. H. Austin}, {Nature (London)} {\bf 358}, {600} {(1992)}.


\bibitem{Reguera_prl}
{D. Reguera, A. Luque, P. S. Burada, G. Schmid, J. M. Rubi, and P. H\"anggi}, {Phys. Rev. Lett.} {\bf 108}, {020604} {(2012)}.


\bibitem{Rubi_prl}
{P. Malgaretti, I. Pagonabarraga, and J. M. Rubi}, {Phys. Rev. Lett.} {\bf 113}, {128301} {(2014)}.


\bibitem{Mukhopadhyay_prl}
{A. K. Mukhopadhyay, B. Liebchen, and P. Schmelcher}, {Phys. Rev. Lett.} {\bf 120}, {218002} {(2018)}.

\bibitem{Sens_prl}
{Q. Vagne and P. Sens}, {Phys. Rev. Lett.} {\bf 120}, {058102} {(2018)}.


\bibitem{Slapik_prl}
{A. Slapik, J. Luczka, P. H\"anggi, and J. Spiechowicz}, {Phys. Rev. Lett.} {\bf 122}, {070602} {(2019)}.

\bibitem{PhysicaA}
{M. F. Carusela, A. J. Fendrik, and L. Romanelli}, {Physica A} {\bf 388}, {4017} {(2009)};
{S. Bouzat}, {Physica A} {\bf 389}, {3933} {(2010)};
{C. Zeng, A. Gong, and Y. Tian}, {Physica A} {\bf 389}, {1971} {(2010)}.

\bibitem{Bader_PNAS}
{J. S. Bader, R. W. Hammond, S. A. Henck, M. W. Deem, G. A. McDermott, J. M. Bustillo, J. W. Simpson, G. T. Mulhern, and J. M. Rothberg}, {Proc. Natl.	Acad. Sci.} {\bf 96}, {13165} {(1999)}.


\bibitem{Bhagat_MBEC}
{A. A. S. Bhagat, H. Bow, H. W. Hou, S. J. Tan, J. Han, and C. T. Lim}, {Med. Biol. Eng. Comput.} {\bf 48}, {999} {(2010)}. 

\bibitem{Aksimentiev_BJ}
{A. Aksimentiev, J. B. Heng, G. Timp, and K. Schulten}, {Biophys. J.} {\bf 87}, {2086} {(2004)}.

\bibitem{Schwartz_Nano}
{M. J. Skaug, L. Wang, Y. F. Ding, and D. K. Schwartz}, {ACS Nano} {\bf 9}, {2148} {(2015)}.

\bibitem{Muller}
{S. Matthias and F. M\"uller}, {Nature (London)} {\bf 424}, {53} {(2003)}.


\bibitem{Newby_R}
{P. C. Bressloff and J. M. Newby}, {Rev. Mod. Phys.} {\bf 85}, {135} {(2013)}.


\bibitem{Burada_CPC} 
{P. S. Burada, P. H\"anggi, F. Marchesoni, G. Schmid, and P. Talkner}, {ChemPhysChem} {\bf 10}, {45} {(2009)}.

\bibitem{Burada_prl}
{D. Reguera, G. Schmid, P. S. Burada, J. M. Rubi, P. Reimann, and P. H\"anggi},  {Phys. Rev. Lett.} {\bf 96},  {130603}  {(2006)}.


\bibitem{Slapik_pra}
{A. Slapik, J. Luczka, and J. Spiechowicz}, {Phys. Rev. Applied} {\bf 12}, {054002} {(2019)}.


\bibitem{Karimi_pre}
{H. Karimi, M. R. Setare, and A. Moradian}, {Phys. Rev. E} {\bf 102}, {012610} {(2020)}. 


\bibitem{Nagaoka_pre}
{B. Lindner, L. Schimansky-Geier, P. Reimann, P. H\"anggi, and M. Nagaoka}, {Phys. Rev. E} {\bf 59}, {1417} {(1999)}.

\bibitem{Borromeo_prl}
{M. Borromeo and F. Marchesoni}, {Phys. Rev. Lett.} {\bf 99}, {150605} {(2007)}.

\bibitem{Suresh}
{S. Suresh}, {Acta Mater.} {\bf 55}, {3989} {(2007)}.


\bibitem{Reguera_pre}
{J. M. Rubi and D. Reguera}, {Chem. Phys.} {\bf 375}, {518} {(2010)}. 

\bibitem{Khatri_JSTAT}
{N. Khatri and P. S. Burada}, {J. Stat. Mech.} {\bf 2021}, {073202} {(2021)}. 


\bibitem{Khatri_pre}
{N. Khatri and P. S. Burada}, {Phys. Rev. E} {\bf 102}, {012137} {(2020)}. 


\bibitem{Khatri_JCP}
{N. Khatri and P. S. Burada}, {J. Chem. Phys.} {\bf 151}, {094103} {(2019)}.



\bibitem{Mahmud_nature}
{G. Mahmud, C. J. Campbell, K. J. Bishop, Y. A. Komarova, O. Chaga, S. Soh, S. Huda, K. Kandere-Grzybowska, and B. A. Grzybowski}, {Nat. Phys.} {\bf 5}, {606}  {(2009)}.

\bibitem{Maier_nlett}
{C. Holz, D. Opitz, J. Mehlich, B. J. Ravoo, and B. Maier}, {Nano Lett.} {\bf 9}, {4553}  {(2009)}.


\bibitem{Cussler}
 {E. L. Cussler}, {Diffusion: Mass Transfer in Fluid Systems} {(Cambridge University Press, 1997)}.

    
\end{thebibliography}
\end{document}